\documentclass[12pt]{article}

\newcommand{\be}{\begin{equation}}
\newcommand{\ee}{\end{equation}}
\newcommand{\bean}{\begin{eqnarray*}}
\newcommand{\eean}{\end{eqnarray*}}

\newcommand{\bea}{\begin{eqnarray}}
\newcommand{\eea}{\end{eqnarray}}

\newcommand{\sgn}{{\rm sgn}}

\newcommand{\const}{{\rm const}}

\newcommand{\p}{\partial}

\title{Membrane fluids and Dirac string fluids}
\author{M. G. Ivanov\thanks{{\tt e-mail: mgi@mi.ras.ru}}\\
{\small\em Moscow Institute of Physics and Technology,}\\
{\small\em  141~700, Dolgoprudnyi, Moscow Region, Russia }}
\date{May 4, 2005}

\begin{document}
\maketitle
\abstract{
  There are two different methods to describe membrane (string)
 fluids, which use different field content.
  The relation between the methods is clarified by
 construction of combined method.

   Dirac membrane field appears naturally in new approach.
   It provides a possibility to consider new aspects
 of electrodynamics-type theories with electric and
 magnetic sources.
   The membrane fluid models automatically prohibit simulatenos
 existence of electric and magnetic currents.

   Possible applications to the dark energy problem are
 mentioned.
}

\section{Introduction}

  Recently various models of membrane (string) fluids are considered.
  These models describe continuous distributions of non-intersecting
 membranes (strings).
  Each space-time point belongs to world surface of one membrane,
 i.e. foliation of space-time is specified (to describe membrane
 fluid one has to specify foliation and density of membrane matter).

  String fluid models are applicable in various realms of physics
 and cosmology. E.g. one can apply zero-brane fluid to describe
 dark energy (on dark energy see e.g. \cite{arefeva} and references
 wherein). Other applications are related to string theory and
 relativistic elasticity theory.

  There are two different approaches to string fluids,
 which have different field contents.
  These approaches are not equivalent.

  Both methods describes fluid of $(n-1)$-dimensional membranes
 (with $n$-dimensional world surfaces $\bf V_\varphi$)
 in $D$-dimensional space-time $\bf M$ by means of $(D-n)$-form $J$
 (membrane field intensity), which
 is Hodge dual of $n$-form current density $j=*^{-1}J$.

  To specify foliation one need form $J$ up to non-zero scalar factor.

  The action, written in terms of $J$ has the similar form for both methods
\be
   S=\int\limits_{\bf M}*L_0(\|J\|),
\ee
 where
\be
   \|J\|=\sqrt{\frac1{(D-n)!}J_{M_1\dots M_{D-n}}\,J^{M_1\dots M_{D-n}}},\\
\ee
  $\|J\|$ represents density of membrane media in space-time $\bf M$.
  Lagrangian $L_0$ is minus energy density in attendant frame.
  (We use $(-,+,+,\dots,+)$ signature of space-time metric.)

   Equation of state is described by relation
\be
  P_{\rm fluid}=L_0-L_0'\,\|J\|.
\label{equation_of_state}
\ee
   Here $P_{\rm fluid}$ is pressure in directions orthogonal
 to membrane world surface.

   Energy-momentum tensor has the following form
\be
   T_{MN}=L_0\,P_{MN}+P_{\rm fluid}\,\bar P_{MN}.
\label{T_MN}
\ee
   $P_{MN}$ is projector to world surface ${\bf V}_\varphi$
 (surface $\varphi=\const$), $\bar P_{MN}=g_{MN}-P_{MN}$
 is projector to directions orthogonal to world surface.

  Due to conservation of membrane current $J$ is closed
\be
  \delta j=0,\qquad\Leftrightarrow\qquad dJ=0.
\label{Jconserv}
\ee
  Form $J$ has to satisfy the following condition
 (orthogonality to membrane world surface ${\bf V}_\varphi$)
\be
  J_{M_1M_2\dots  M_{D-n}}\,A^{M_1}=0,\qquad
  \forall~A\in T{\bf V}_\varphi.
\label{Jsurf}
\ee

  Two methods differ by parametrization of $J$ in terms of dynamical field.

  The form $J$ is closed, so
 the most obvious way is to consider membrane field
 in terms of $(n-1)$-form potential $I$
 (similar to non-linear electrodynamics)
\be
  J_I=dI.
\ee
  In this approach the action has the following form
\be
   S_I[I]=\int\limits_{\bf M}*L_0(\|J_I\|),
\label{S_I}
\ee
  This method was introduced independently by several authors,
 see e.g. \cite{spall,gibbons,roberts} and references wherein.
  Conservation condition (\ref{Jconserv}) is satisfied automatically.
  To satisfy condition (\ref{Jsurf}) one has to restrict himself by certain
 class of solutions (it can be done by introduction of
 additional term to action, which impose appropriate constraint).
  These solutions are generalizations of potential flow in
 standard hydrodynamics.

  Potential $I$ could be considered as membrane field intensity
 for Dirac membrane (string) fluid (see section \ref{Dirac}).

  Another way was initially suggested (in the particular
 case of constant volume form $\Omega_{\bf F}$)
 to describe single membranes \cite{hos1}--\cite{bf2}
 (papers \cite{bf1,bf2} suggest dual interpretation, which
 correspond to use of $*J$ form instead of $J$).
  It was later geometrized and generalized to membrane fluids and other
 continuous media in \cite{MGI-DAN}--\cite{MGI-med}.
  Using this method one has to consider auxiliary $(D-n)$-dimensional
 space $\bf F$ (points $\phi\in\bf F$ numerate membranes ${\bf V}_\phi$)
 with volume form, which
 represents density of membrane media in auxiliary space $\bf F$,
\be
  \Omega_{\bf F}=f(\phi)\,d\phi^1\wedge\dots\wedge d\phi^{D-n},
\ee
 where $\phi^\alpha$, $\alpha=1,\dots,D-n$ are coordinates at $\bf F$,
 and mapping
\bea
  &&\varphi:{\bf M}\to{\bf F},\\
\nonumber
  &&\varphi(X)=(\varphi^\alpha(X)).
\eea
  Membrane field intensity is defined by the following relation
\be
  J_\varphi=\varphi^*\Omega_{\bf F}
   =f(\varphi)\,d\varphi^1\wedge\dots\wedge d\varphi^{D-n}.
\ee
  In this approach the action has the following form
\be
   S_\varphi[\varphi]=\int\limits_{\bf M}*L_0(\|J_\varphi\|).
\label{S_phi}
\ee

  Conservation condition (\ref{Jconserv}) is satisfied automatically.
  Moreover, condition (\ref{Jsurf}) is also satisfied automatically.
  Every solution of model (\ref{S_I}),
 which satisfies condition (\ref{Jsurf}),
 is also solution (potential flow)
 of model (\ref{S_phi}).
  Field equations of model (\ref{S_I}) are field equations
 of model (\ref{S_phi}) with extra condition of flow potentiality.
  These potential flows admit also dual description with $J$ replaced
 by $*K=*\frac{\p L_0}{\p J}=L_0'\frac{*J}{\|J\|}$.

  The theory written in terms of fields $\varphi$ has no linear
 dependence in equations of motion.
  So, the theory admits straightforward Hamiltonian formulation and
 formal quantization \cite{MGI2}.
  The other approach to quantize the theory is based upon the
 Nambu brackets (see \cite{zachos} and references wherein).

  Here we present a method, which involves both $\varphi$ and $I$.
  It allows to reveal new symmetries in electrodynamics-type theories
 and membrane fluid models.

\section{Free membrane fluids}

   To involve constraint
\be
   J_I=J_\varphi\qquad\Leftrightarrow\qquad dI=\varphi^*\Omega_{\bf F}
\ee
 one can introduce the following action
\be
  S_{\varphi I}[\varphi,I,K]
  =\int\limits_{\bf M}\left\lbrace *L_0\left(\sqrt{(J_I,J_\varphi)}\right)
   +\frac12(J_\varphi-J_I)\wedge*K\right\rbrace.
\label{action-FIK}
\ee
  Here
\be
  (J_I,J_\varphi)
  =\frac1{(D-n)!}(J_{I})_{M_1\dots M_{D-n}}\>
              (J_{\varphi})^{M_1\dots M_{D-n}}.
\ee

   Field equations are
\be
  J_I=J_\varphi
\ee
  (so, we use notation $J=J_I=J_\varphi$),
\bea
  &&\delta\left(K-L_0'\frac{J}{\|J\|}\right)=0,\\
  &&(J_{(\alpha)},\delta K)=0,
\eea
 here $\delta=*^{-1}d*$,
$$
  J_{(\alpha)}=(-1)^{\alpha+1}\,f(\varphi)\,d\varphi^1
  \wedge\dots\widehat{d\varphi^\alpha}\dots\wedge d\varphi^{D-n}.
$$
  The term $\widehat{d\varphi^\alpha}$ has to be skipped.

  If one exclude fields $I$ and $K$ from field equation,
 then the equations appear to be equivalent to field
 equations for action (\ref{S_phi}).
  So, actions (\ref{S_phi}) and (\ref{action-FIK}) describe
 the same system in different ways.

  According to \cite{MGI-del} field $K$ could be considered
 as analogue of momentum in mechanics
 (in this case $J$ is analogue of velocity).

  $K$ and $I$ admits gauge-like transformations
\bea
\nonumber
   K&\to&K+\delta\lambda,\\
   I&\to&I+d\mu.
\label{gauge-like1}
\eea

\section{Interactions of membrane fluids}

  To introduce interaction to action (\ref{action-FIK}) one could
add to action extra terms.
  To preserve the symmetries of the theory the terms have to be
written in terms of fields $J_\varphi$ and $I$ only.

  For arbitrary action
\be
  S[\varphi]=\int\limits_{\bf M}L(J_\varphi),
\ee
 which involves fields $\varphi^\alpha$
 through $J_\varphi$ only, the equaction of motion, produced
 by variation by $\varphi^\alpha$, is of the following form
\be
 \left(J_{(\alpha)},\delta\left[\frac{\p L}{\p J_\varphi}\right]\right)=0.
\ee

   If the function $L(J_\varphi)$ is written in the form
 $L(J_\varphi)=L_1(\|J_\varphi\|,J_\varphi)$,
 then field equation acquire the form
\bea
 &&\left(J_{(\alpha)},
 \delta\left[
 \frac{\p L_1}{\p\|J_\varphi\|}\frac{J_\varphi}{\|J_\varphi\|}
 \right]
 +{\cal F}\right)=0,\\
 &&{\cal F}=\delta\left[\frac{\p L_1}{\p J_\varphi}\right]
\eea
 (If action is written in terms of $J_\varphi$, $I$ and $K$,
 the equations of motion have the similar form.)

   The field ${\cal F}$, the equations of motion involve,
 satisfies the condition
\be
  \delta{\cal F}=0.
\ee

  In section \ref{electric} field ${\cal F}=\pm*F$, so
$\delta{\cal F}=0\quad\Leftrightarrow\quad dF=0$.
 I.e. ``electric'' membrane fluid interacts with the field,
which admits no ``magnetic'' currents.

  In section \ref{magnetic} field ${\cal F}=\pm F$, so
$\delta{\cal F}=0\quad\Leftrightarrow\quad \delta F=0$.
 I.e. ``magnetic'' membrane fluid interacts with the field,
which admits no ``electric'' currents.

\section{Membrane fluid as electric-type source\label{electric}}

  One can easily introduce interaction of membrane fluid
 with closed form $F$
 to action (\ref{action-FIK}) by inserting standard
 interaction term
\be
  S_{\rm el.}[\varphi,I,K,A]
  =\int\limits_{\bf M}\left\lbrace* L_0(\sqrt{(J_I,J_\varphi)})
   +\frac12(J_\varphi-J_I)\wedge*K-\frac12F\wedge*F
   +(-1)^{n+1}A\wedge J_I\right\rbrace,
\label{action-el1}
\ee
 where $F=dA$ is closed $(n+1)$-form. So, $dF=0$.
  Interaction term could be written in other equivalent form
\bean
  (-1)^{n+1}\int\limits_{\bf M} A\wedge J_I
 =(-1)^{n+1}\int\limits_{\bf M} A\wedge dI
 =\int\limits_{\bf M} [-d(A\wedge I)+dA\wedge I]=\\
 =-\int\limits_{\p\bf M} A\wedge I
  +\int\limits_{\bf M} dA\wedge I.
\eean
  Up to surface term
\be
  S_{\rm el.}[\varphi,I,K,A]
  =\int\limits_{\bf M}\left\lbrace*L_0(\sqrt{(J_I,J_\varphi)})
   +\frac12(J_\varphi-J_I)\wedge*K
   -\frac12F\wedge*F+F\wedge I\right\rbrace,
\label{action-el2}
\ee

   Field equations are
\be
  J_I=J_\varphi
\ee
  (we use notation $J=J_I=J_\varphi$),
\bea
\label{eomA0}
  && \delta F=*^{-1}J,\\
\label{eomI0}
  &&\delta\left(K-L_0'\frac{J}{\|J\|}\right)+2\,\sgn(g)(-1)^{D-n-1}*F=0,\\
  &&\left(J_{(\alpha)},
    \delta\left(L_0'\frac{J}{\|J\|}\right)+\sgn(g)(-1)^{D-n}*F\right)=0.
\eea
  $K$ and $I$ still admits gauge-like transformations (\ref{gauge-like1}).

  One could derive from field equation (\ref{eomI0}), which
 is produced by variation of action by $I$, the condition
$$
  dF=0.
$$
  Consequently the action (\ref{action-el1}), (\ref{action-el2})
 does not allow any method of introduction of magnetic charges, which
 does not involve field $I$.

\section{Membrane fluid as magnetic-type source\label{magnetic}}

  To introduce interaction of magnetic type with field
 of form $F$ one can use method similar to Dirac strings.
  This method is partially similar to dual variable electrodynamics
 (see \cite{mandelstam, volovich} and references wherein).

  The appropriate action has the form
\be
  S_{\rm mag.}[\varphi,I,K,A]
  =\int\limits_{\bf M}\left\lbrace* L_0(\sqrt{(J_I,J_\varphi)})
   +\frac12(J_\varphi-J_I)\wedge*K-\frac12F\wedge*F\right\rbrace,
\label{action-mag1}
\ee
 where $(D-n-1)$-form $F$ is defined by
\be
   F=dA+I.
\ee
  Here field $I$ represents Dirac string (membrane) distribution.
\be
  dF=J_I.
\ee

   Field equations are
\be
  J_I=J_\varphi
\ee
  (we use notation $J=J_I=J_\varphi$),
\bea
\label{eomA}
  && \delta F=0,\\
\label{eomI}
  &&\delta\left(K-L_0'\frac{J}{\|J\|}\right)+2(-1)^{D-n}F=0,\\
  &&\left(J_{(\alpha)},
    \delta\left(L_0'\frac{J}{\|J\|}\right)+(-1)^{D-n-1}F\right)=0.
\eea
  $K$ still admits gauge-like transformations (\ref{gauge-like1}).
  Gauge-like transformation of $I$ now involves potential $A$
\bea
\nonumber
  I&\to&I+d\mu,\\
  A&\to&A-\mu.
\label{gauge-like2}
\eea
   This transformation ({\em generalised gradient transformation})
 generalizes standard gauge transformation
 for field $A$.
   It corresponds to deformation of Dirac strings (membranes)
 with fixed boundaries.

   By transformation (\ref{gauge-like2}) one can set $A=0$ (i.e. $F=I$).
   Moreover, one can set $A=0$ {\em before} variation of action.
\be
  S_{\rm mag.}[\varphi,I,K]
  =\int\limits_{\bf M}\left\lbrace* L_0(\sqrt{(J_I,J_\varphi)})
   +\frac12(J_\varphi-J_I)\wedge*K-\frac12I\wedge*I\right\rbrace,
\label{action-mag2}
\ee
   It is possible because field equation (\ref{eomA}), which
 is produced by variation of action by $A$,
 could be derived from field equation (\ref{eomI}), which
 is produced by variation of action by $I$.

  Consequently the action (\ref{action-mag1}), (\ref{action-mag2})
 does not allow any method of introduction of electric charges, which
 does not involve field $I$.

\section{Dirac strings (membranes)\label{Dirac}}

  This section just formulate standard Dirac string approach
to introducing of magnetic charges to electrodynamics-type theories
in the context of the paper.
Theories considered below have quadratic Lagrangians, but the
approach is also applicable to non-linear theories e.g. to Born-Infeld
model.

  We wrote the word ``string'' with quotation marks to
emphasize that it is actually $q$-dimensional membrane.

  The standard electrodynamics-type action has the following form
\be
  S[A(x)]=-\int\limits_{\bf M}
  \left\lbrace\frac12F\wedge*F+A\wedge*j_e\right\rbrace,
\label{action1}
\ee
 where
\be
  F=dA
\label{F=dA}
\ee
 is $(q+1)$-form field intensity,
 ${\bf M}$ is $D$-dimensional space-time region,
 $A$ is $q$-form potential,
 $j_e$ is $q$-form ``electric'' current density.

 External derivative of equation (\ref{F=dA}) produce ``first pair''
of Maxwell-type equations
\be
  dF=0.
\label{1st1}
\ee
  Variation of action (\ref{action1}) by $A$ produce ``second pair''
of Maxwell-type equations
\be
  \delta F=(-1)^qj_e.
\label{2nd}
\ee

  Current $j_e$ is conserved
\be
  \delta j_e=0\qquad \Leftrightarrow\qquad d*j_e=0,
\label{conserve1}
\ee
 so action (\ref{action1}) is invariant
under gradient transformations \be
  A\longrightarrow A+df,
\label{transform1}
\ee
where $f$ is arbitrary $(q-1)$-form.

  To introduce ``magnetic'' currents one introduces Dirac ``strings'' $\bf I$
(open membranes with $(q+1)$-dimensional world surfaces $\bf I$,
$q$-dimensional boundaries of membranes $\p\bf I$ are world
surfaces of magnetic sources) and consider action (\ref{action1})
with integration in region $\bf M\backslash I$ with cut-off $\bf
I$ instead of original region $\bf M$.
  At world surface $\bf I$ one has continuity condition
for field intensity $F$
\be
  \exists\lim_{x\to x_0\in{\bf I\backslash\p I}} F(x).
\label{FatBcontinuity}
\ee

   Dirac strings are similar to infinitely thin tube, which
transfer magnetic flux from monopole to infinity or to other
monopole of opposite sign.
   It allows to replace cut-off $\bf I$
and condition (\ref{FatBcontinuity}) by
redefinition of field $F$
\be
  F=dA+I,
\label{F=dA+B}
\ee
where $I$ compensates magnetic flux inside tube.

  New ``first pair'' of Maxwell-type equations is
\be
  dF=dI\qquad\Leftrightarrow\qquad \delta*^{-1}F=*^{-1}dI=j_m.
\label{1st2}
\ee
  Here magnetic current
\be
  j_m=*^{-1}dI
\label{j_m}
\ee
is defined.
  By definition (compare with (\ref{conservm1}))
\be
  \delta j_m=*^{-1}d^2I=0.
\label{conservm2}
\ee

  One can use the old form of action (\ref{action1}) with new
definition of $F$ (\ref{F=dA+B}) and reproduce the same ``second pair''
of Maxwell-type equations (\ref{2nd}) by variation of potential $A$.

If world surface $\bf I$ is defined by the following system
\bea
\nonumber
  b^\alpha(x)|_{\bf I}&=&0,\qquad \alpha=1,\dots,q+1,\\
  b^0|_{\bf I}(x)&\geq&0,
\eea
and describes ``magnetic'' charge $Q_m$
(numeration of functions $b^\alpha$ has to be in
agreement with surface orientation), then
\be
   I=Q_m\theta(b^0(x))\>d\theta(b^1(x))\wedge\dots\wedge d\theta(b^{q+1}(x)),
\label{string2field}
\ee
where $\theta$ is Heaviside theta-function.

   Dirac ``string'' describes singular (like point particles)
magnetic currents with world
lines (surfaces) $\p\bf I$.
   Magnetic currents are conserved due to the following property
of $\p$ operation
\be
 \p^2{\bf I}=0\qquad\Leftrightarrow\qquad d^2I=0.
\label{conservm1}
\ee
   Dirac ``string'' has to go through points with zero $j_e$
(to avoid interaction of magnetic flux inside tube with electric currents).
To guarantee the possibility to find such surface one could require
electric currents to be singular (like point particles) too.

   Dirac ``string'' world surface is uniquely defined by $(q+1)$-form $I$,
so actually we do not need Dirac ``string'' itself, but form $I$
(Dirac ``string'' density) only.
   This point of view is more flexible, one has no longer
 obligation to restrict himself by fields $I$ of form (\ref{string2field}).
   It reveal more general symmetry like (\ref{gauge-like2}).

   Form $J_I=dI$ is membrane field intensity.
   It satisfies conditions (\ref{Jconserv}) and (\ref{Jsurf}).

   Form $I$ is not unique.
   It is defined up to gradient transformation (\ref{gauge-like2}).
   By appropriate choice of gauge form $I$ could satisfy
 condition (\ref{Jsurf}) (with ${\bf V}_\varphi$ replaced
 by world surface $\bf I$ of Dirac membrane), but not (\ref{Jconserv}).
   So form $I$ is similar to non-conserved membrane
 field intensity. It describes {\em Dirac membrane fluid}.
   It is natural to replace condition (\ref{Jsurf}), for
 field $I$, which holds only in special gauge by more
 subtle condition
\be
  {dI}_{M_1M_2\dots  M_{D-n}}\,A^{M_1}=0,\qquad
  \forall~A\in T{\bf V}_\varphi.
\label{Bsurf}
\ee

\section{Conclusion}

\subsection{Models variety}
  The approach proposed here is applicable not only to
introduce interaction between membrane fluid and electrodynamics-type
field.
  Instead of membrane fluid one can use membrane elastic media
(see \cite{MGI-med}), which also involve membrane density
form $J_\varphi$.
  E.g. one could construct different classical models of elastic
electrons.

  The form of action of electrodynamics-type field is not fixed too.
 The method is also applicable for nonlinear models like
 Born-Infeld-type models, etc.

  Even membrane fluid models are able to reproduce wide range
of state equations. State equation $p(\rho)$ is related with
Lagrangian $L_0(\|J\|)=-\rho(\|J\|)$ by the following relation:
\be
 \ln\|J\|=\int\frac{d\rho}{\rho+p(\rho)}.
\ee

  E.g. state equation
\be
  p=w\rho
\ee
 is reproduced by action
\be
  S_w[\varphi]=-\int\limits_{\bf M}*\|J_\varphi\|^{1+w},
\quad {\rm or}\quad
  S_w[\varphi,I,K]
  =\int\limits_{\bf M}\left\lbrace *(J_I,J_\varphi)^{\frac{1+w}2}
   +\frac12(J_\varphi-J_I)\wedge*K\right\rbrace.
\ee
  $\rho=-L_0=\|J_\varphi\|^{1+w}$,
 see equations (\ref{equation_of_state}), (\ref{T_MN}).

   Zero-brane fluid with negative pressure $-1.5<w<-0.5$
 could be useful to describe dark energy \cite{arefeva}.

\subsection{No-classical monopole hypothesis}

  The approach described has an unexpected heuristic issue.
  One can conclude that simultaneous existence of
 electric and magnetic currents is not possible.
  The electric-type sources considered in section \ref{electric}
 automatically suggests non-existence of magnetic type source
 (as consequence of field equations).
  Similarly, the magnetic-type sources considered in section \ref{magnetic}
 automatically suggests non-existence of electric type source
 (as consequence of field equations).

  This difficulties have simple physical interpretation.
  The Dirac string approach require the possibility to
 draw Dirac string through the regions with no electric charges.
  If charges are point-like, then one able to draw Dirac strings
 in any situation.
  This requirement is crucial, because Dirac strings are
 similar to magnetic flux tubes. If a point electric charge intersect
 Dirac string it has to suffer infinitely large impact of
 Lorentzian force. Fortunately in classical theory with point
 particles the probability of this process is zero.

  Nevertheless the classical theory with point particle
 is not self-consistent.
  If one replace point electric charges
 by continuous distributions, then there are configurations
 of charges, which make impossible to draw Dirac strings.
  Similarly, if magnetic charges are delocalized, then
 instead of single Dirac strings one has Dirac string distribution,
 which occupies finite volume.

  In quantum theory one could expect delocalization of classical singular
 charges due to uncertainty principle.
  So, infinite impact with infinitesimal probability could produce
 the finite effect of interaction of electrical charges with
 Dirac strings.

  The arguments above suggests the hypothesis, that there is
 no self-consistent classical theory with action, which involves both
 electric charges and magnetic monopoles.

\subsection*{Acknowledgement}
  The author is grateful to I.V. Volovich and I.Ya. Aref'eva.
  The work was partially supported by grants RFFI 02-01-01084,
 NSh-1542.2003.1, and by joint grant Y2-PM-11-08 of CRDF and RF Ministry of Education and Science.

\end{document}